\documentclass{PoS}

\usepackage{dcolumn}
\usepackage{bm}

\usepackage[square,numbers,compress,sort]{natbib}
\usepackage{graphicx}
\usepackage{amssymb,amsmath}
\usepackage{multirow}
\usepackage{slashed,xfrac}
\usepackage[makeroom]{cancel}

\usepackage{epsfig,psfrag,rotating,soul}
\usepackage{rotfloat}
\usepackage[normalem]{ulem}

\graphicspath{{figs/}}

\psfrag{lk}{$l_k$}
\psfrag{lm}{$l_m$}
\psfrag{blk}{$\bar {l}_k$}
\psfrag{blm}{$\bar {l}_m$}

\def\ne{\boldsymbol{n}_e}
\def\nm{\boldsymbol{n}_\mu}
\def\nt{\boldsymbol{n}_\tau}
\def\mode{|\ne|}
\def\modmu{|\nm|}
\def\modtau{|\nt|}
\def\tme{\theta_{\mu e}}
\def\tte{\theta_{\tau e}}
\def\ttm{\theta_{\tau \mu}}
\def\cme{c_{\mu e}}
\def\cte{c_{\tau e}}
\def\ctm{c_{\tau \mu}} 

\def\modelname{ISS-$\cancel{\rm LFV}\hskip-.1cm_{\mu e}$}
\def\rotationO{\mathcal O}

\definecolor{vdrgreen}{rgb}{0.0, 0.7, 0.0}

\def\be{\begin{equation}}
	\def\ee{\end{equation}}
\def\bea{\begin{eqnarray}}
	\def\eea{\end{eqnarray}}

\title{Charged lepton flavour violation from low scale seesaw neutrinos \thanks{IFT-UAM/CSIC-17-038, FTUAM-17-6}}

\ShortTitle{Charged lepton flavour violation from low scale seesaw neutrinos}

\author{Valentina De Romeri \\
        Departamento de F\'{\i}sica Te\'orica and Instituto de F\'{\i}sica Te\'orica, IFT-UAM/CSIC,\\
Universidad Aut\'onoma de Madrid, Cantoblanco, 28049 Madrid, Spain\\
        \email{valentina.deromeri@uam.es}}
        
        \author{Maria Jose Herrero \\
        Departamento de F\'{\i}sica Te\'orica and Instituto de F\'{\i}sica Te\'orica, IFT-UAM/CSIC,\\
Universidad Aut\'onoma de Madrid, Cantoblanco, 28049 Madrid, Spain\\
        \email{maria.herrero@uam.es}}
        
         \author{Xabier Marcano \\
       Departamento de F\'{\i}sica Te\'orica and Instituto de F\'{\i}sica Te\'orica, IFT-UAM/CSIC,\\
Universidad Aut\'onoma de Madrid, Cantoblanco, 28049 Madrid, Spain\\
        \email{xabier.marcano@uam.es}}
\author{\speaker{Francesca Scarcella}%
        \\
        Departamento de F\'{\i}sica Te\'orica and Instituto de F\'{\i}sica Te\'orica, IFT-UAM/CSIC,\\
Universidad Aut\'onoma de Madrid, Cantoblanco, 28049 Madrid, Spain \\ and Dipartimento di Fisica e Astronomia,\\ Universit\`{a} di Bologna, via Irnerio 46, 40126 Bologna, Italy\\
        \email{francesca.scarcella@studio.unibo.it}}

\abstract{In the work presented here, we have studied the impact of right handed neutrinos, which are introduced to account for the evidence of neutrino masses, on charged lepton flavour violating observables. In particular, we have focused on the loop induced decays of the Z boson into two leptons of different flavour. We have performed a numerical study of the rates predicted for these processes within the Inverse Seesaw model, specifically considering scenarios where $ \mu -e $ transitions are suppressed. Our conclusion, after comparison with the most relevant experimental constraints, is that branching ratios as large as $ 10^{-7} $ can be predicted in the $ \tau -\mu $ or $ \tau -e $ channels, together with heavy neutrinos having masses of the TeV order. Such rates could be accessible at next generation colliders.}

\FullConference{Corfu Summer Institute 2016 "School and Workshops on Elementary Particle Physics and Gravity"\\
                 31 August - 23 September, 2016\\
                 Corfu, Greece}

\begin{document}

\section{Introduction}

Flavour violating transitions are strictly prohibited by the Standard Model (SM) of particle physics in its leptonic sector, as a direct consequence of the neutrinos being assumed to be massless. 
This prediction has been proven wrong by experiments: a neutrino produced in a specific flavour state, either $ \nu_e, \nu_\mu, \nu_\tau $, has an oscillating probability to be detected after some time as a different flavour state. Neutrino oscillations are explained through misalignment of the interaction and mass basis and thus interpreted as proof that neutrinos are massive, although extremely light. 

On the other hand, no process involving Lepton Flavour Violation (LFV) has been observed between charged leptons, despite an intense experimental program currently searching for it. Much effort is put into these searches because any observation of such transitions would indicate physics beyond the SM, and give us hints to understand the laws of Nature at a more fundamental level. Some of the current upper bounds on charged LFV (cLFV) transitions are summarised in Tables \ref{LFVsearch} and ~\ref{LFVsearchII}.
The most stringent limits come from the present bounds on LFV muon decays, from the MEG and SINDRUM  experiments. These have constrained the branching ratios for LFV $\mu$ decays and $\mu$ to $e$ conversions in nuclei to be below $10^{-12} $-$ 10^{-13}$  at 90\% C.L..  LFV decays of the Higgs and Z boson are also being searched for at LHC, where ATLAS and CMS have both established interesting bounds, summarised in Table~\ref{LFVsearchII}. The sensitivities to these decays are expected to improve significantly, both at LHC and at experiments that are under study at present, such as linear colliders or a future circular collider. In particular, the expected sensitivity to LFV Z decays (LFVZD) at  future linear colliders has been estimated as $10^{-9}$, while a Future Circular  $e^+ e^-$ Collider (such as FCC-ee/TLEP), is further estimated to be able to produce up to $10^{13}$ $Z$ bosons, meaning that the sensitivities to LFVZD rates could be improved up to roughly $\sim 10^{-13}$. LFVZD decays have been chosen as the main subject of the study we are presenting here, Ref.~\cite{DeRomeri:2016gum}.

\begin{table}[t!]
\begin{center}
\begin{tabular}{|c|c|c|}
\hline
LFV Observable & Present Bound  $(90\%CL)$ & Future Sensitivity \\
\hline
BR$(\mu\to e\gamma)$ &  $4.2\times10^{-13}$ (MEG 2016) & $4\times 10^{-14} $ (MEG-II\\
BR$(\tau\to e\gamma)$ & $3.3\times10^{-8}$ (BABAR 2010)~ & $ 10^{-9}$ (BELLE-II)\\
BR$(\tau\to \mu\gamma)$ & $4.4\times10^{-8}$ (BABAR 2010) & $ 10^{-9}$ (BELLE-II)\\
BR$(\mu\to eee)$ &   $1.0\times10^{-12}$ (SINDRUM 1988) & $10^{-16}$ Mu3E (PSI) \\
BR$(\tau\to eee)$ & $2.7\times10^{-8}$ (BELLE 2010)& $10^{-9,-10}$ (BELLE-II)\\
BR$(\tau\to \mu\mu\mu)$ & $2.1\times10^{-8}$ (BELLE 2010) & $10^{-9,-10}$ (BELLE-II)\\
BR$(\tau\to \mu\eta)$ & $2.3\times10^{-8}$ (BELLE 2010) & $10^{-9,-10}$ (BELLE-II)\\
CR$(\mu-e,{\rm Au})$ & $7.0\times10^{-13}$ (SINDRUM II 2006)& \\
CR$(\mu-e,{\rm Ti})$ &  $4.3\times10^{-12}$ (SINDRUM II 2004)&$10^{-18}$ PRISM (J-PARC)\\
CR$(\mu-e,{\rm Al})$ &&$3.1\times10^{-15}$ COMET-I (J-PARC)\\
&&$2.6\times10^{-17}$ COMET-II (J-PARC) \\
&&$2.5\times10^{-17}$ Mu2E (Fermilab) \\
\hline
\end{tabular}
\caption{ Present upper bounds and future expected sensitivities for cLFV transitions. }\label{LFVsearch}
\vskip .5cm

\begin{tabular}{|c|c|c|c|}
\hline
LFV Observable & Present Bound  $(95\%CL)$ \\
\hline
BR$(Z\to\mu e)$ & $1.7\times10^{-6}$ (LEP 1995), $7.5\times10^{-7}$ (ATLAS 2014)\\
BR$(Z\to\tau e)$ & $9.8\times10^{-6}$ (LEP 1995) \\
BR$(Z\to\tau\mu )$ &$1.2\times10^{-5}$ (LEP 1995), $1.69\times10^{-5}$ (ATLAS 2014)\\
BR$(H\to\mu e)$ & $3.6\times10^{-3}$ (CMS 2015)\\
BR$(H\to\tau e)$ & $1.04\times10^{-2}$ (ATLAS 2016), $0.7\times10^{-2}$  (CMS 2015) \\
BR$(H\to\tau\mu )$ & $1.43\times10^{-2}$ (ATLAS 2016), $1.51\times10^{-2}$ (CMS 2015) \\
\hline
\end{tabular}
\caption{Present upper bounds at $95\%$ CL on LFV decays of $Z$ and $H$ bosons.}\label{LFVsearchII}
\end{center}
\end{table}

The search for cLFV plays an important role in neutrino physics. Neutrino masses necessarily induce loop level LFV in the charged lepton sector, with model dependent rates. This provides a very valuable discriminant between the many different models able to account for neutrino masses. Amongst these, the ones generally considered to be most attractive  belong to the class of so-called Seesaw models. These relate the lightness of neutrinos to the existence of a very high energy scale where some kind of new physics comes into play. The extreme lightness of neutrinos is explained through the small ratio between the vacuum expectation value of the Higgs and the high scale of new physics.
The most natural realisations of the Seesaw mechanism are those which assume the existence of neutrinos of right handed (RH) chirality: neutrinos are the only fundamental particles that appear in the SM exclusively trough their left handed (LH) chiral component. In this case the new physics scale is the mass of the RH neutrinos. As an example, consider the Type-I Seesaw model (SS-I) \cite{Minkowski:1977sc}, the simplest of the family. The SS-I extends the SM with RH neutrinos introducing both a Dirac and a Majorana mass term:
\begin{equation}
 \mathcal{L}^{SS-I}= -  Y^{ij}_\nu \overline{L_i} \widetilde{\phi} \nu_{Rj} + \frac{1}{2} M_{R}^{ij} \overline{\nu_{Ri}^C} \nu_{Rj} + h.c.\,,
\end{equation}
where ${\lbrace L_i\rbrace}_{1=1,2,3}$ are the 3 SM lepton doublets, $\Phi$ is the SM Higgs doublet, $\widetilde{\Phi}=\imath \sigma_2 \Phi^*$.
The Dirac mass term is originated through EWSB from the Yukawa interaction: $m_D=Y_\nu \langle \Phi\rangle $.
The limit  $m_D\ll M_R$ is then taken. This is physically justified by the fact that the Majorana mass term is invariant under the full SM gauge group, so that it does not need to be related to the EWSB scale. 
Defining a vector of RH fields (we omit the flavour indices for shortness) $\nu=(\nu_L^C, \nu_R)^T$ , the mass Lagrangian can be written as a Majorana mass term:
\begin{equation}
\mathcal{L}_M= -\frac{1}{2} \overline{\nu^C} M \nu + h.c. = -\frac{1}{2} \overline{\nu^C} \left(\begin{array}{c c} 0 & m_D\\ m_D^T & M_R\end{array}\right) \nu + h.c. \,.
\end{equation}
 Since the matrix $M $ is symmetrical, it can be brought in diagonal form by a unitary transformation $U^{\nu}$:
\begin{equation}
diag(m_{n_1},..., m_{n_9})= {U^{\nu}}^{T}M U^{\nu}\,,
\end{equation}
Upon diagonalisation we obtain the physical mass eigenstates. These are Majorana neutrinos, two for each generation.
In the one generation case, the masses of the two neutrinos in the Seesaw limit are:
\be
m_{n_1}\simeq \frac{m_D^2}{M_R}\,,\quad \quad m_{n_2}\simeq M_R\,.
\label{ss1}
\ee 
One of the mass eigenstates represents a beyond the SM heavy  neutrino of mass approximately equal to $ M_R $, while the other has its mass suppressed by the same heavy scale\footnote{The light mass eigenstate $n_1$ is predominantly composed of $\nu_L$, while the heavy eigenstate $n_2$ is mainly composed of $\nu_R$. The admixture of the singlet neutrino state $\nu_R$ in $n_1$ and that of the usual neutrinos $\nu_L$ in $n_2$ are of the order of
$m_D/M_R$. }.
The same mechanism operates in the three generations case, yielding three very heavy exotic neutrinos states and three light ones, that we identify with the neutrinos we have observed.

With $\mathcal{O}(1)$ Yukawa couplings (of the order of the top Yukawa), light neutrino masses at the $eV$ scale are obtained by taking $M_R\sim10^{15}\;\mathrm{GeV}$. For Yukawas of the order of the electron Yukawa, $\mathcal{O}(10^{-6})$, the heavy mass scale must be set at $M_R\sim1\;\mathrm{TeV}$.

As we mentioned, the presence of massive neutrinos will induce, at loop level, cLFV processes. In the Type-I Seesaw model, however, their amplitudes turn out to be suppressed by powers of the same small ratio $ m_D/M_R$ that suppresses neutrino masses, making them inaccessible in any reasonable future experimental scenario. This is not, however, a characteristic common to all Seesaw models: some of them, the so called low scale Seesaw models, combine a new physics scale of the $ \mathcal{O} (TeV) $ with $\mathcal{O}(1)$ Yukawa couplings. 

\section{The Inverse Seesaw Model}

We have considered in this work one of the most known low scale Seesaw models, the Inverse Seesaw (ISS) \cite{Bernabeu:1987gr}. The ISS extends the SM particle content by introducing RH neutrinos in three pairs of opposite lepton number (LN): $\nu_{R}$ (with LN $L_{\nu_{R}}\equiv 1$) and $ X $ ($L_{X}\equiv-1$). The SM is extended with the Lagrangian:
\begin{equation}
{\mathcal{L}}_{ISS}\equiv
-{Y_{\nu}}^{ij}\overline{L_{i}}\widetilde{\Phi}\nu_{Rj}
-{M_{R}}^{ij}\overline{\nu_{Ri}^{C}}X_{j} -
\dfrac{1}{2}{\mu_{X}}^{ij}\overline{X_{i}^{C}}X_{j} +h.c.,\,
\end{equation}
where $\mu_X$ is
a Majorana complex  $3\times 3$ symmetric mass matrix. This last term is the only one, in the total Lagrangian, that violates LN conservation: the mass parameter $\mu_{X}$ is then protected by the correspondent symmetry. Based on this argument, the Lagrangian is studied in the Seesaw limit $\mu_X \ll m_D \ll M_R$.
The mass Lagrangian after EWSB is:
\begin{equation}
{\mathcal{L}}_{ISS}^{M}=-\dfrac{1}{2}
\begin{pmatrix}
\overline{\nu_{L}} & \overline{\nu_{R}^{c}} &  \overline{X^{c}}
\end{pmatrix} 
\begin{pmatrix}
0 & m_{D} & 0 \\ 
m_{D}^{T} & 0 & M_{R} \\ 
0 & M_{R}^{T} & \mu_{x}
\end{pmatrix} 
\begin{pmatrix}
\nu_{L}^{c} \\ 
\nu_{R} \\ 
X
\end{pmatrix} + h.c.
\equiv -\dfrac{1}{2} 
\overline{\nu^{c}}~M_{ISS} ~\nu + h.c. \,.
\end{equation}
In analogy with the Type I Seesaw case, we can find a basis of Majorana neutrino fields in which the mass matrix is diagonal. In the one SM generation case we obtain one light state and two heavy ones, with masses:
\begin{align}
 m_1 & \approx \frac{m_{D}^2}{M_{R}^2} \mu_X\, \quad \quad m_{2,3}\approx  \pm M_{R} + \frac{\mu_X}{2}.
\end{align}
The light neutrino mass $m_1$ is proportional to $\mu_X$, so that it becomes massless in the $\mu_X \rightarrow 0$ limit. The other two masses $m_{2}, m_{3}$ are close to each other, with their splitting also being proportional to the small scale of LN violation.
A similar behaviour occurs in the case of three SM generations and three generations of extra fermionic singlet pairs: we obtain three light Majorana neutrinos and six heavy ones. In this case, the mass matrix $M_{\mathrm{ISS}}$ can be diagonalised by blocks,
leading, at first order in $ m^2_D/M^2_R$, to the following $3 \times 3$ light neutrino mass matrix:
\begin{equation}\label{Mlight}
M_{\mathrm{light}} \simeq m_D (M_R^T)^{-1} \mu_X M_R^{-1} m_D^T\,.
\end{equation}
This may then be diagonalised to find the three light mass eigenstates that are to be identified with the light neutrinos observed in oscillation experiments. The six heavy neutrinos form three pseudo-Dirac pairs, their masses approximately given by the eigenvalues of $M_R$ with splitting of order $\mathcal O(\mu_X)$.

The introduction of extra RH neutrinos produces a departure from the SM interaction Lagrangian. Such departure is due to: (i) the presence of a Yukawa interaction term for the neutrinos; (ii) the non-alignment of the neutrino physical and interaction basis. With respect to the SM, we obtain extra terms coupling the Higgs to the leptons and modified terms describing the weak interaction. We list below the relevant interaction Lagrangians for the ISS in the physical basis with three generations and a total of nine Majorana neutrinos, given in the Feynman - t'Hooft gauge. The presence of the fermionic singlets induces flavour to be violated by mixings in both charged and neutral current interactions.
\begin{align}
\mathcal L_W&=-\dfrac g{\sqrt{2}} \sum_{i=1}^{3}\sum_{j=1}^{9} W^-_\mu \bar\ell_i B_{\ell_i n_j} \gamma^\mu P_L n_j + h.c., \\
\label{interactions}
\mathcal L_Z &= -\dfrac g{4c_W} \sum_{i,j=1}^{9}Z_\mu\, \bar n_i \gamma^\mu \Big[C_{n_i n_j} P_L - C_{n_in_j}^* P_R\Big]n_j ,\\
\mathcal L_H &=-\dfrac g{2m_W}  \sum_{i,j=1}^{9} H\,\bar n_i C_{n_in_j}\Big[m_{n_i}P_L+m_{n_j} P_R\Big]n_j,\\
\mathcal{L}_{G^{\pm}} &= -\frac{g}{\sqrt{2} m_W}\sum_{i=1}^{3}\sum_{j=1}^{9} G^{-}\bar{\ell_i} B_{\ell_i n_j} \Big[m_{\ell_i} P_L - m_{n_j} P_R \Big]n_j  + h.c\,,\\ 
\mathcal{L}_{G^{0}} & =-\dfrac {ig}{2 m_W} \sum_{i,j=1}^{9}G^0\, \bar n_i  C_{n_in_j} \Big[m_{n_i}  P_L - m_{n_j} P_R\Big]n_j .  
\end{align}
In the equations above, $P_{L,R}=(1\mp \gamma^5)/2$  are the usual chirality projectors, $g$ is the weak coupling constant, $c^2_W=m^2_W/m^2_Z$, and $m_{n_i}$,  $m_{l_i}$  are the physical
neutrino and lepton masses respectively. 
Finally, $B$ and $C$ are $3\times 9 $ and 
$9\times 9$ dimensional matrices, respectively, 
which are defined as
\begin{equation}
B_{l_in_j}\ \equiv \sum_{k=1}^{3} V^l_{ik} U^{\nu\ast}_{kj}\quad
\mbox{and}\quad
C_{n_in_j}\ \equiv\ \sum_{k=1}^{3}\ U^\nu_{ki}U^{\nu\ast}_{kj}\,,
\end{equation}\label{eq:BCmatrices}
where $U^{\nu}$ is the unitary matrix diagonalizing the full mass matrix. The $3\times 3$ matrix $V^l$ allows for non-alignment of the mass and interaction basis for the charged leptons. If we assume, as we will here, these to be equivalent, then the $B$ matrix simply defines the composition of the interacting LH neutrinos in terms of the physical ones. In particular, the $ 3\times3 $ sub-block of ${B_{l_in_j}}$ obtained for $  i,j=\lbrace 1,2,3 \rbrace$ , encodes the rotation from the interacting LH neutral leptons to the light physical neutrinos. It is an important prediction of the model since it may be directly confronted with experimental data from neutrino oscillations.

\begin{figure}[t!]
\begin{center}
\includegraphics[width=\textwidth]{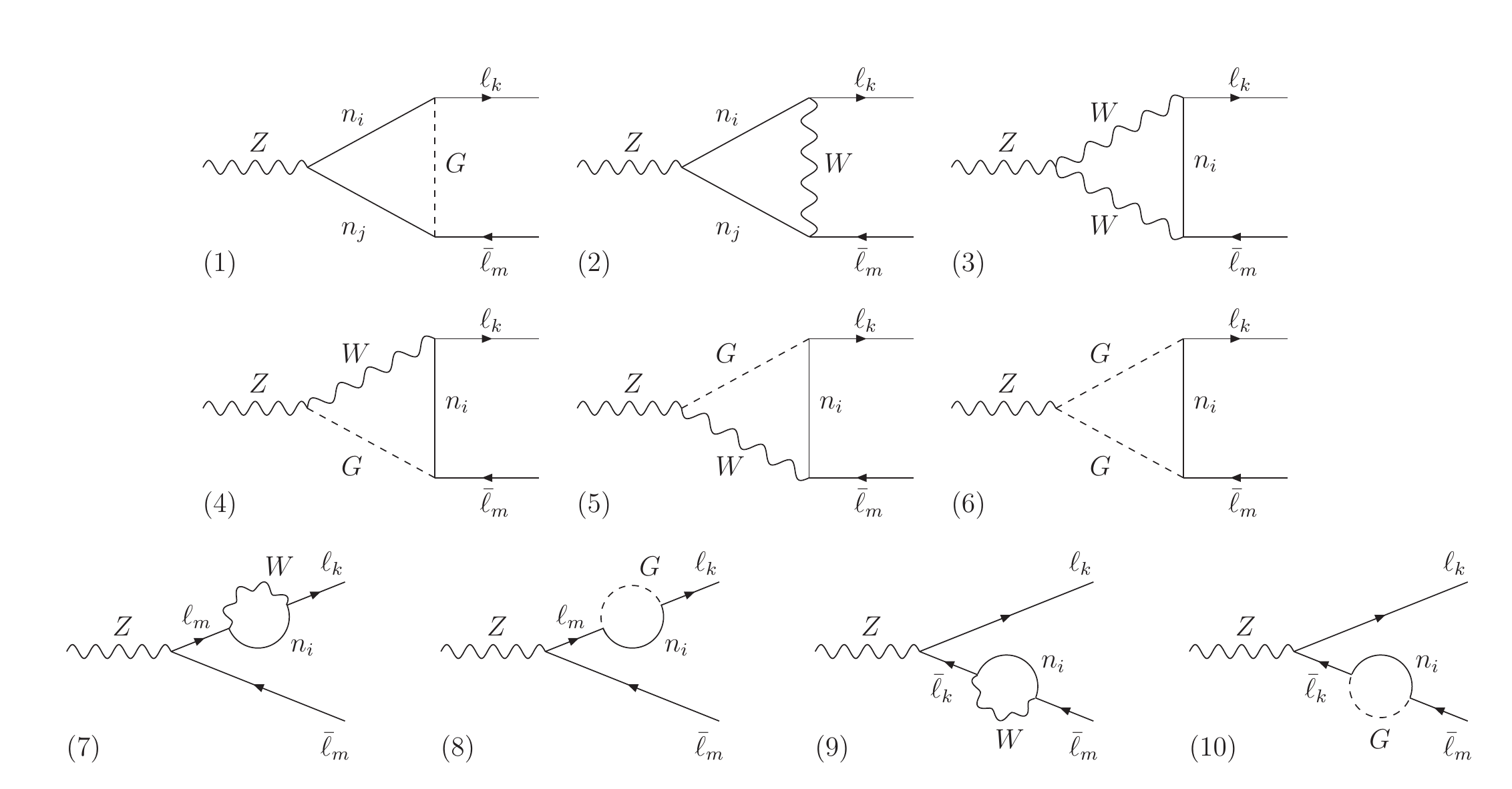}
\caption{One-loop diagrams in the Feynman-t'Hooft gauge for LFV Z decays.}\label{ZDiagrams}
\end{center}
\end{figure}

Notice in particular that the addition of massive neutrinos also allows flavour violating $Z \nu_i\nu_j$ interactions (flavour-changing neutral currents), see Eq. (\ref{interactions}). Together with the charged-current couplings, these interactions will induce, at one-loop level, an effective charged lepton flavour violating vertex $Z \ell_1 \ell_2$ with $ \ell_1 \neq \ell_2 $, see Figure \ref{ZDiagrams}.

\section{The ISS with suppressed $\mu$-$e$ transitions }
\label{Scenarios}

LFVZD processes in the presence of low-scale heavy neutrinos have been studied in recent works, considering the full one-loop contributions \cite{Abada:2014cca} or computing the relevant Wilson coeffcients \cite{Abada:2015zea}. In these works, maximum allowed rates for these decays were found when considering a minimal $3+1$ toy model: this predicted BR($Z\to\tau\mu$) up to $ \mathcal{O}(10^{-8}) $, in the reach of future linear colliders, for a heavy neutrino mass in the few TeV range. When considering more realistic scenarios, such as the (2,3) or (3,3) realisations of the ISS model, smaller rates $\lesssim \mathcal{O}(10^{-9}) $ were achieved. These would be below the sensitivities of future linear colliders and might be accesible only at future circular $e^+ e^-$ colliders. The above results have been  obtained through scans of the parameter space. However, we have shown in Ref.~\cite{DeRomeri:2016gum} that larger maximum allowed rates for BR($Z\to\tau\mu$)  and BR($Z\to\tau e$) can be obtained by considering the particular scenarios we present in this section. This analysis is based on the parametrisation  proposed in Ref.~\cite{Arganda:2014dta} for the ISS model, in which the Majorana mass $ \mu_X $ is computed as an appropriate function of  $ M_R $ and $ Y_\nu $ (cfr. Eq. (\ref{Mlight})), in order to impose a priori agreement with oscillation experiments data:

\begin{equation} \label{muxparametrization}
\mu_X = M_R^T {m_D}^{-1} {M_{\mathrm{light}}}^{exp} ~({m_D^T})^{-1} M_R\,,
\end{equation}

where: 
 \begin{equation} 
 {M^{exp}_{\mathrm{light}}}={U^*_{\rm PMNS}} \mathrm{diag}(m_{\nu_1}\,, m_{\nu_2}\,, m_{\nu_3}) {U^\dagger_{\rm PMNS}}\,.
 \end{equation} 
 The advantage of this procedure lies in the fact that, apart from its important role in suppressing light neutrino masses, $\mu_X $ has little impact on the LFV and RH neutrino induced phenomenology of the model, which is instead controlled mainly by the Yukawa coupling and by the heavy mass scale $ M_R $. Thus, by maintaining these as input parameters, we obtain direct control on the  phenomenological predictions we are interested in, as we now discuss.
 
We have allowed only real values for the $Y_\nu$ matrix, in order to avoid potential constraints from lepton electric dipole moments. For simplicity, we have worked in the basis where the  matrix $ M_R$ is diagonal and assumed its eigenvalues to be degenerate {\it i.e}, $M_{R_{1,2,3}}\equiv M_R$. 

Following the geometrical interpretation discussed in \cite{Arganda:2014dta}, the entries of the real $ 3\times3 $ Yukawa matrix can be interpreted as of the components of three vectors ($\ne,\nm,\nt$):
\begin{equation}
Y_\nu \equiv f 
 \left(\begin{array}{c} \ne \\ \nm \\ \nt \end{array}\right).
\end{equation}
In the definition above we have factored out a positive factor $\emph{f}$. The other parameters that determine the $Y_\nu$ matrix are: three moduli $\mode,\modmu,\modtau$, three relative  flavour angles  $\tme,\tte,\ttm$ and, finally, three extra angles $\theta_1,\theta_2,\theta_3$ that parametrize a global rotation $\rotationO$. We take $\mode,\modmu,\modtau$ of $ \sim O(1) $, so that $\emph{f}$ defines the strength of the coupling. With the assumptions we mentioned above, the only further input parameters of the model are the heavy mass scale $M_R$, which roughly corresponds to the mass of the heavy neutrinos, and the mass of the lightest neutrino. The latter has little influence on the model's predictions; we have set it to $0.01$ eV in our computations, 

The Yukawa matrix enters in the amplitude rates for one-loop induced cLFV processes through the combination $Y_\nu Y_\nu^T$ \cite{Arganda:2014dta}.

 \begin{equation}\label{YukawaCombination}
Y_\nu Y_\nu^T= f^2 \left(\begin{array}{ccc}
 \mode^2 & \ne\cdot\nm   & \ne\cdot\nt \\
 \ne\cdot\nm & \modmu^2 &  \nm\cdot\nt\\
 \ne\cdot\nt &   \nm\cdot\nt  & \modtau^2\end{array}\right),
 \end{equation}
 
so that, for instance, the dominant contribution to $ {\rm BR}(\ell_m\to \ell_k\gamma) $ depends on the combination $\boldsymbol{n}_m\cdot\boldsymbol{n}_k$. This suggests that a selective suppression of the $ {\rm BR}(\ell_\mu \to \ell_e\gamma) $, and of $\mu$-$e$ transitions in general, may be achieved by enforcing the orthogonality of $\ne$ and $\nm$. 
This condition can be easily imposed on the Yukawa couplings. The most general Yukawa matrix that satisfies $\cme=0$ (with $c_{ij}\equiv $ cos $ \theta_{ij}$) is:

\begin{equation}\label{YukawaAmatrix}
Y_\nu=A\cdot \rotationO \quad{\rm with}\quad 
A\equiv f \left(\begin{array}{ccc} \mode & 0 & 0 \\ 0 & \modmu & 0 \\ \modtau \cte & \modtau \ctm & \modtau\sqrt{1-\cte^2-\ctm^2}\end{array}\right),
\end{equation}

where we have factored out the global rotation matrix $\rotationO$, which does not affect the phenomenological predictions.

\begin{figure}[h!]
\begin{center}
\includegraphics[width=.65\textwidth]{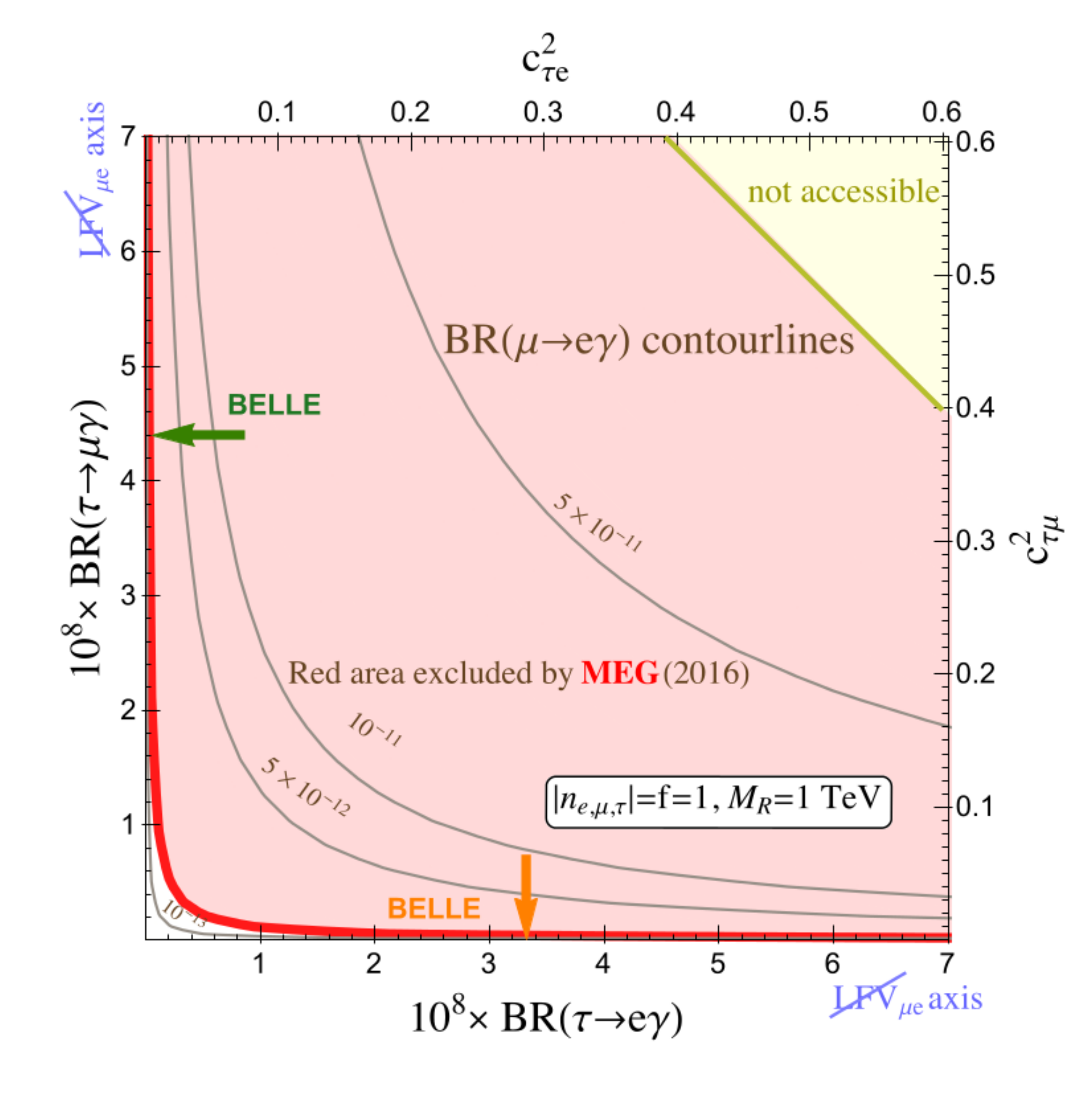}
\caption{Contour lines for BR($\mu\to e\gamma$) as function of BR($\tau\to e\gamma$) and BR($\tau\to \mu\gamma$) rates, for $M_R=1$~TeV, $|\boldsymbol{n}_{e,\mu,\tau}|=f=1$ . The yellow area represents the region that cannot be accessed with real Yukawa matrices. The red area is excluded by upper bound on $\mu\to e\gamma$ from MEG, while the orange (green) arrow marks the present upper bound on $\tau\to e\gamma$ ($\tau\to\mu\gamma$) from Babar.}\label{fig:radiativecosine}
\end{center}
\end{figure}

While imposing orthogonality of $\ne$ and $\nm$ sets to zero the main contribution to LFV one loop processes in the $\mu$-$e $ sector, subdominant contributions exist: the  $\mu$-$e $ transition can occur through the combination of $\tau$-$\mu $ and $\tau$-$e$ transitions. This is encoded in terms which have the form $Y_\nu Y_\nu^T Y_\nu Y_\nu^T$ and are proportional to $\ctm^2$ and $\cte^2$.
Consider for instance the radiative decay $ \mu \to e \gamma $, which is particularly relevant due to the strong bound recently set by MEG. The numerical significance of the subdominant contribution to this decay amplitude is illustrated in Figure \ref{fig:radiativecosine}. Here we consider a Yukawa matrix of the form given in Eq. (\ref{YukawaAmatrix}), and show BR($\mu\to e\gamma$), computed through the full one-loop expressions, as a function of $\ctm^2$ and $\cte^2$. Predictions for BR($\tau \to \mu \gamma$) and BR($\tau \to e \gamma$) are represented for comparison in the vertical and horizontal axes, respectively\footnote{ BR($\tau \to \mu \gamma$) has been verified to be largely independent of $\cte$ (and vice versa for $\tau \to e \gamma$).}.

We notice that, despite the predictions for BR($\mu\to e\gamma$) being orders of magnitude smaller than those for the $\tau$ radiative decays, the MEG bound on this process still excludes the sector of parameter space where BR($\tau\to e \gamma$) and BR($\tau\to\mu\gamma$) are simultaneously significant. Large amplitudes for cLFV processes in either of the  $ \tau $ sectors, are allowed only along the axes of the plot (defined by $\cte=0$ and $\ctm=0$ respectively).

Motivated by the above observations, we have defined the following two classes of scenarios: the TM class ($\cte=\cme=0$), composed by scenarios that induce relevant cLFV phenomenology in the $ \tau$-$\mu $ sector, but always give negligible rates for ${\rm LFV}_{\mu e}$ and ${\rm LFV}_{\tau e}$ transitions; and the TE class ($\ctm=\cme=0$) which may lead to large rates only in the $\tau$-$e$ sector. These two classes compose what we have called the \modelname.

In Table \ref{TMscenarios} we list some examples of scenarios that belong to the TM class. We will use them for numerical estimations in section  \ref{max Z decays}. Equivalent examples for the TE scenarios are obtained by exchanging $\mu$ and $e$ in the TM ones.  While we will refer in the following to the TM class mainly, corresponding results are obtained considering the TE class.

\begin{table}[h!]
\begin{center}

\begin{tabular}{|c|c|c|c|c|l|}
\hline
 Scenario Name& $\ctm$ & $\mode$ & $\modmu$ & $\modtau$ 	\\
\hline
TM-1 & $1/\sqrt2$ & 1 & 1 & 1 \\
TM-2 & $1$ & 1 & 1 & 1 \\
TM-3 & $1/\sqrt2$ & 0.1 & 1 & 1 \\
TM-4 & $1$ & 0.1 & 1 & 1 \\
TM-5 & $1$ & $\sqrt2 $& $1.7$ &$ \sqrt3 $\\
TM-6 & $1/3$ & $\sqrt2 $& $\sqrt3$ & $\sqrt3 $\\
TM-7 & $0.1$ & $\sqrt2 $& $\sqrt3 $& $1.1 $\\
TM-8 & $1$ & $1/2$ & $1/3$ & $1/4$\\
 \hline
\end{tabular}
\caption{Examples of Yukawa coupling matrices defining scenarios of the TM class. We have used these scenarios for the numerical estimates presented in section \ref{max Z decays}. Equivalent scenarios of the TE class are easily obtained by exchanging $\mu$ and $e$. }\label{TMscenarios}
\end{center}
\end{table}

\section{LFV Z decay rates }\label{Z decays}

\begin{figure}[t]
\begin{center}
\includegraphics[width=.48\textwidth]{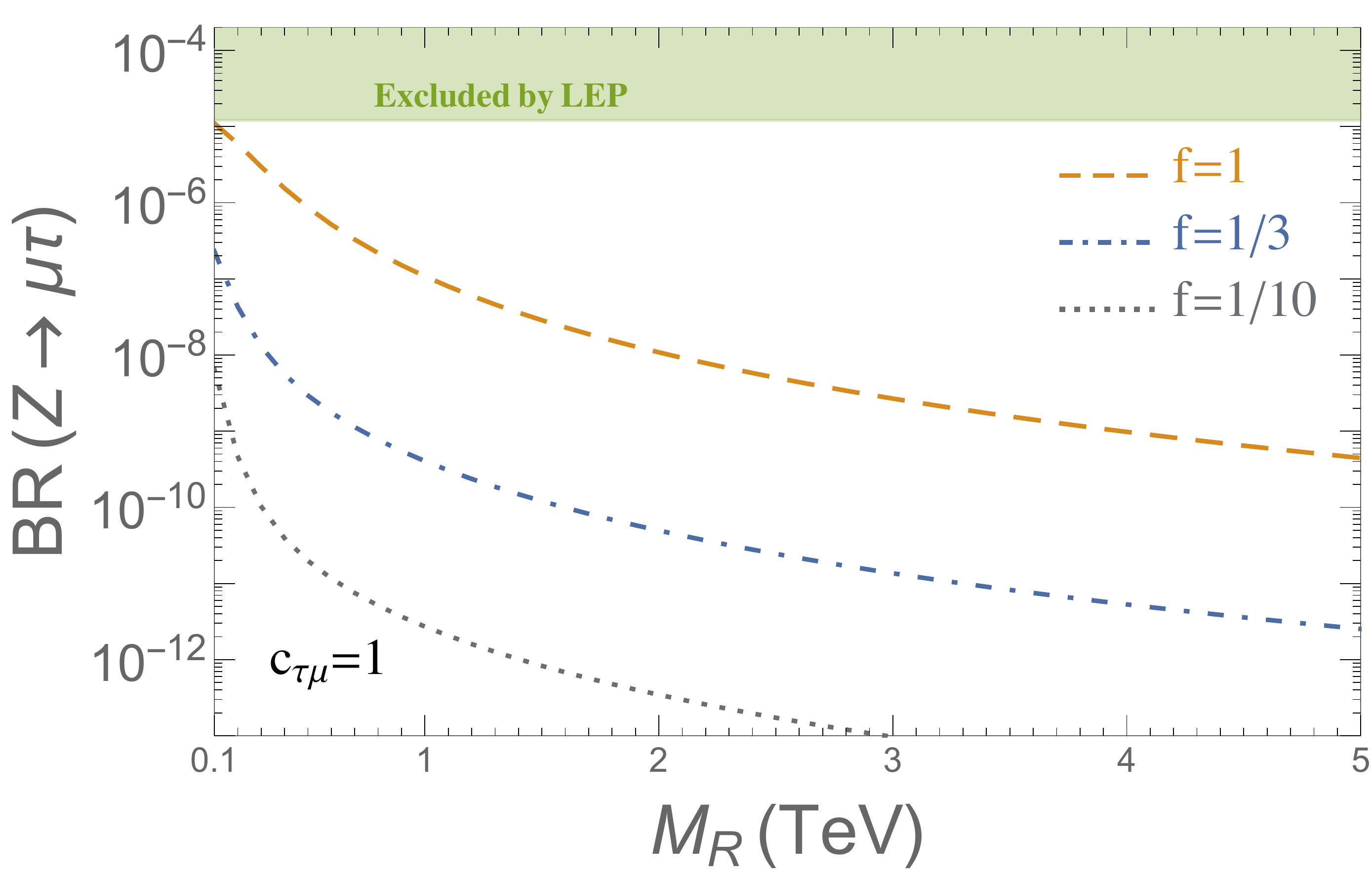}\, \,
\includegraphics[width=.48\textwidth]{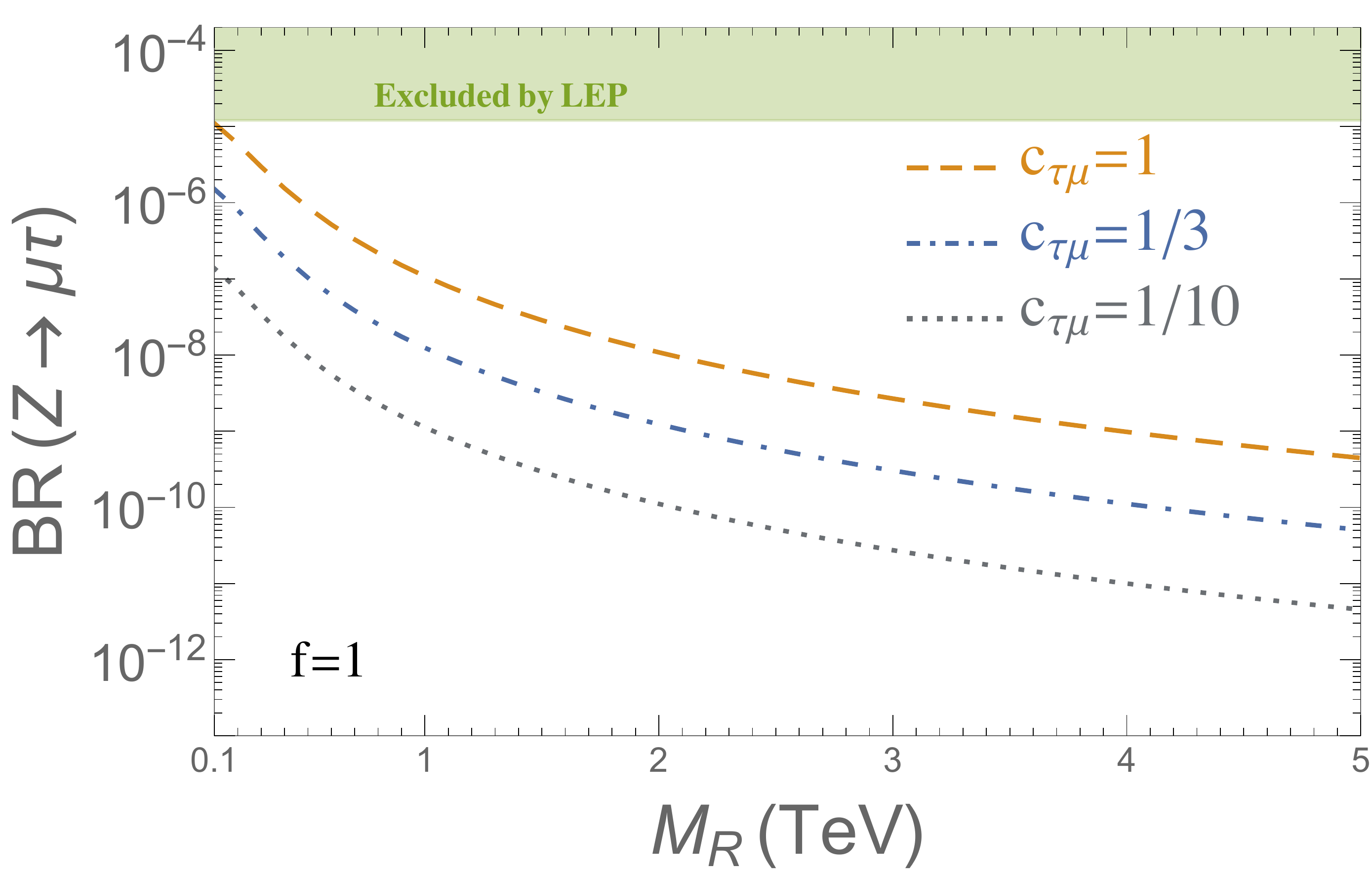}
\includegraphics[width=.48\textwidth]{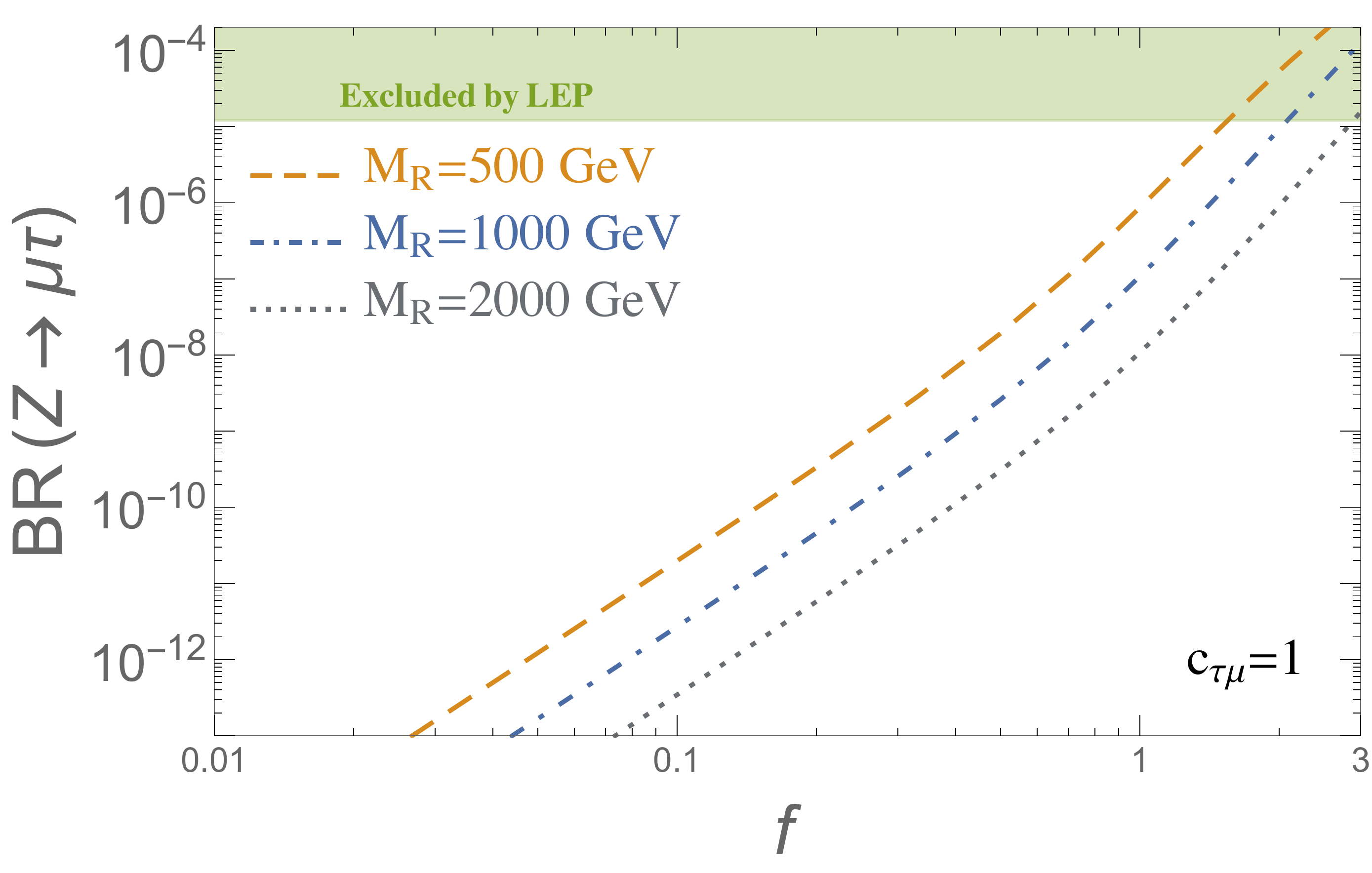}~~
\includegraphics[width=.48\textwidth]{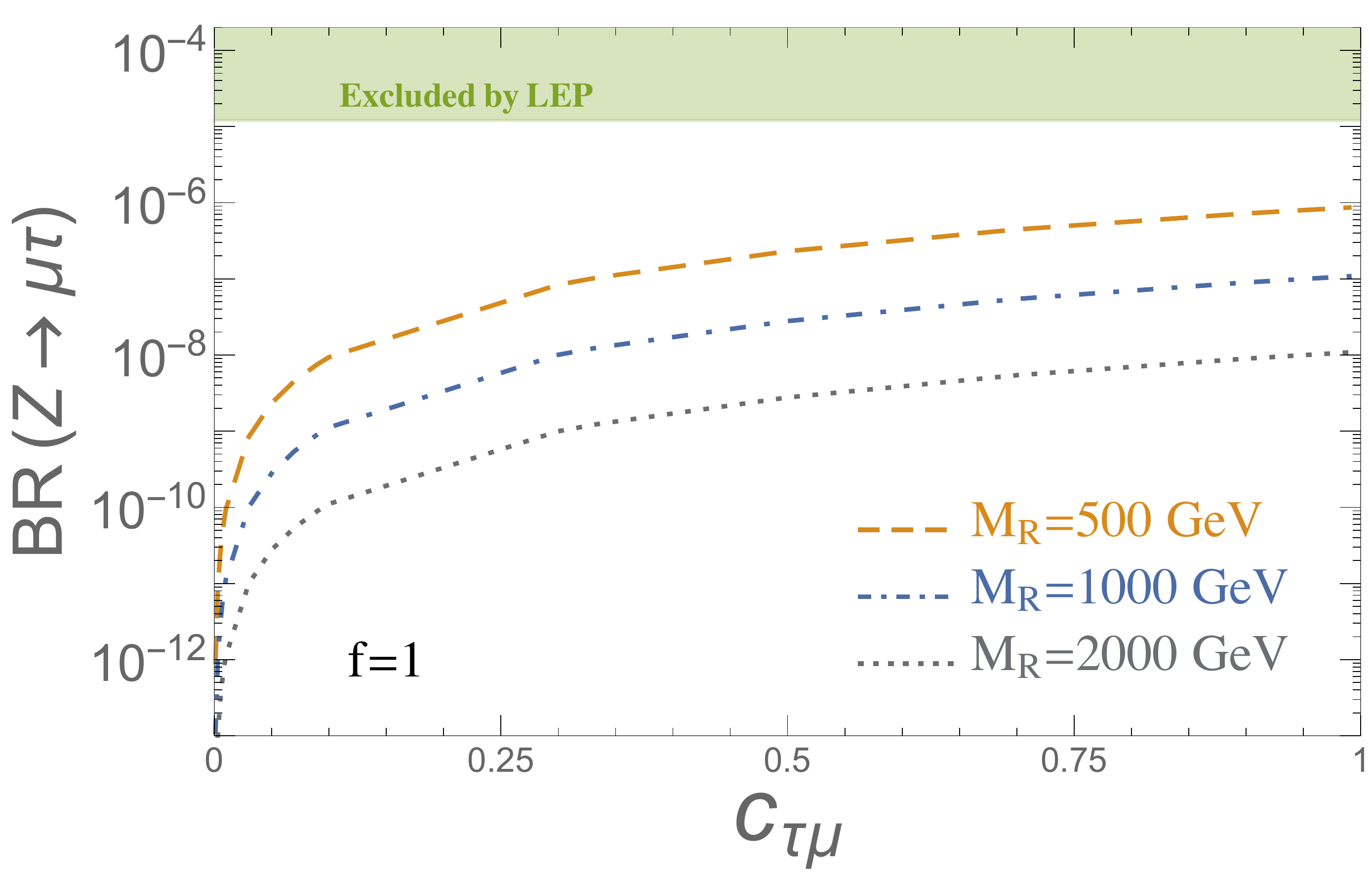}
\caption{Predictions for BR($Z\to\tau\mu$) within the ISS model as function of the heavy neutrino mass parameter $M_R$ (two upper panels), the neutrino Yukawa coupling strength $f$ (lower left panel) and $\ctm$ (lower right panel) for various choices of the relevant parameters. In all plots we have fixed, $\cte=0$ and $|\boldsymbol{n}_{e,\mu,\tau}|=1$.  
The upper shadowed areas are excluded by LEP. Similar results for BR($Z\to\tau e$) by exchanging $\cte$ and $\ctm$.
}\label{Ztaumu}
\end{center}
\end{figure}

The TM realisations of the ISS model can result in large rates for the LFV $Z$ decay $Z\to\tau \mu$. We have verified this by computing with our Mathematica code the full one loop rates for different values of the free parameters $M_R$, $f$, $\modtau$, $\modmu $ $\mode $ and $\ctm$. The rates decrease with the heavy scale $M_R$ (for large values of this mass scale the heavy neutrinos decouple from the light ones) and grow with the Yukawa coupling strength $f$. For large values of $ f $ and small values of $ M_R $ the predicted rates are close to the upper bound from LEP and to present LHC sensitivity. We display in Figure~\ref{Ztaumu} the behaviour of the BR($Z\to\tau\mu$) rates with the $M_R$, $f$ and $\ctm$ parameters for fixed values of $\mode=\modmu=\modtau=1$, $\cte=0$ and $\rotationO=\mathbb I$.

\section{Constraints}

Before concluding on the maximum allowed branching ratio for LFVZD within the ISS-$\cancel{\rm LFV}\hskip-.1cm_{\mu e}$, we must find out which area of parameter space is allowed by the current experimental observations. We have evaluated numerically with our Mathematica code the model's predictions for those observables which are expected to provide the most stringent constraints. All bounds have been applied at the 3$\sigma$ level. Full details of our computations can be found in Ref.~\cite{DeRomeri:2016gum}.

\subsection{LFV lepton decays  }

By construction, the \modelname ~scenarios that we are studying suppress LFV in two out of the three $\ell_i$-$\ell_j$ sectors. The only relevant LFV decays within the TM class occur in the $\tau$-$\mu$ sector. The most important experimental bound are set by $\tau\to\mu\gamma$ and $\tau\to\mu\mu\mu$ decays.
We have computed the full one loop rates and compared them with their experimental upper limits from Babar and Belle, respectively (see Table \ref{LFVsearch}). 
Note that for high values of $f$ and small $M_R$, rates in the range of the exclusion bounds are obtained.
 
\begin{figure}[t!]
\begin{center}
\includegraphics[width=0.49\textwidth]{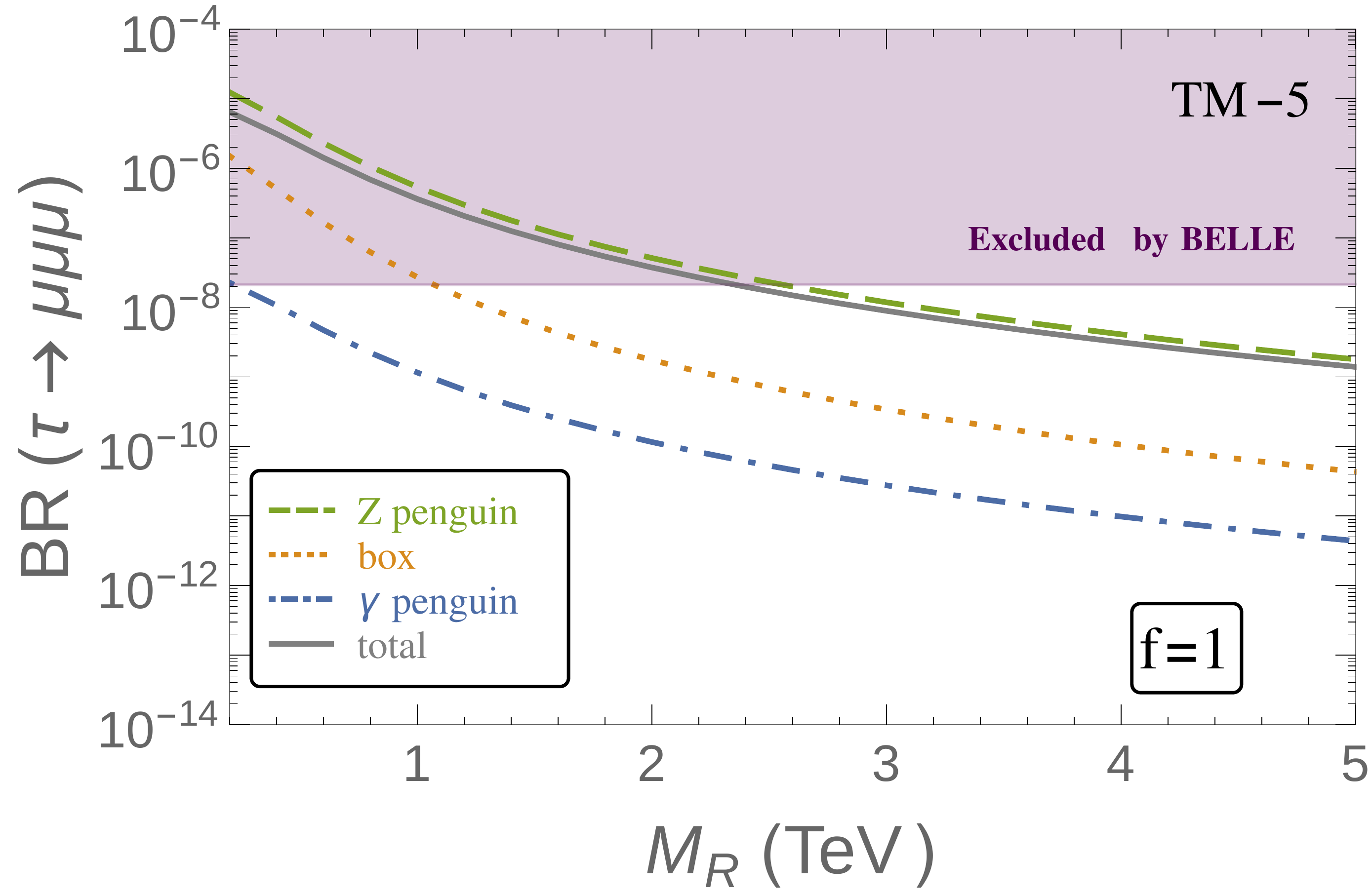}
\includegraphics[width=.49\textwidth]{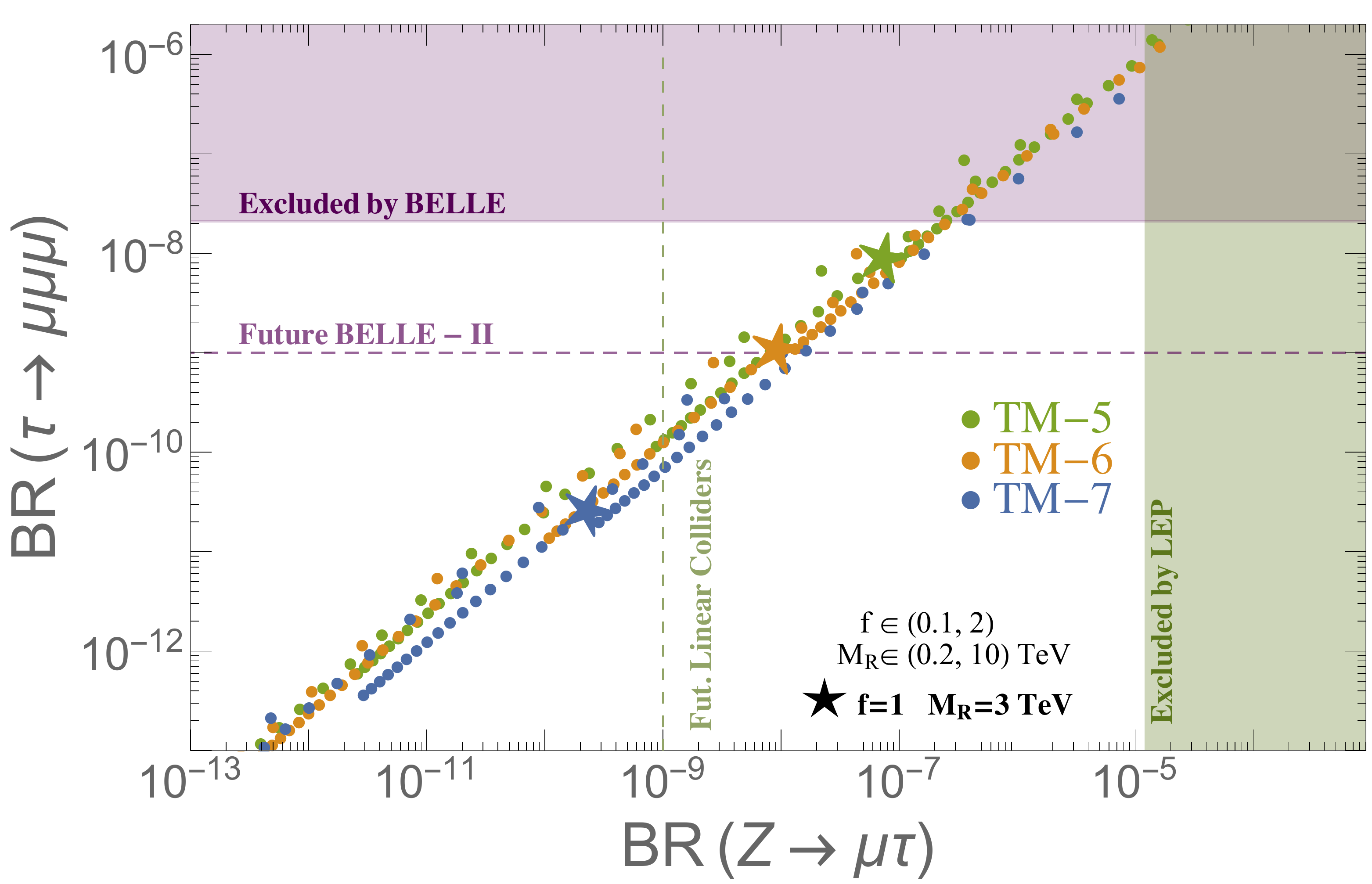}
\caption{Left panel: BR($\tau\to\mu\mu\mu$) as function of $M_R$ for $f=1$ in the TM-5 scenario. The full prediction is decomposed in its contributions, showing that the Z penguin is the dominant one. Right panel: correlation plot for BR($Z\to\mu\tau$) and BR($\tau\to\mu\mu\mu$) for scenarios TM-5 (green), TM-6 (yellow) and TM-7 (blue) defined in Table \ref{TMscenarios}. The dots are obtained by varying $f\in(0.1,2)$ and $M_R\in(0.2,10)$~TeV and the stars are for the reference point $f=1$ and $M_R=3$~TeV.
}\label{tau3mu}
\end{center}
\end{figure}

It is interesting to compare the predictions for the LFV $Z$ decays with those for the three body LFV lepton decays. The BR($\tau\to\mu\mu\mu$) rates are strongly dominated by the $Z$ penguin contributions. This implies an important correlation between $\tau\to\mu\mu\mu$ and $Z\to\tau\mu$, since the effective vertex responsible of the latter appears in the Z penguin diagram. 
In fact, it turns out that the whole three body decay amplitude can be approximated fairly well by considering the  $Z$ penguin diagram alone, as shown in Figure \ref{tau3mu}. 
Notice how the present upper bound on $\tau\to\mu\mu\mu$ from Belle alone excludes rates of BR($Z\to\tau\mu)$ higher than $\sim 2 \times 10^{-7}$. 

\subsection{Lepton Flavour Universality}

It has been shown by previous studies that leptonic and semileptonic decays of pseudoscalar mesons can put important constraints on the mixing between the active and the sterile neutrinos in the ISS. In particular, the most severe bounds arise from the violation of lepton universality in leptonic kaon decays. We computed $\Delta r_k$  and compared it with the experimental results from the NA62 collaboration.
We found that the resulting exclusions become more important as the ratio between $\mode$ and $\modmu$ departs from one. In general however the constraints obtained from this observable are weaker than those obtained from LFV lepton decays.

\subsection{The invisible decay width of the Z boson}

The presence of sterile neutrinos affects the tree level predictions of the $Z$ invisible width, since they modify the couplings of the active neutrinos to the $Z$ boson.  When extra heavy sterile neutrinos are added to the SM, the $Z$ invisible decay width will turn out to be smaller than its SM value. Comparing our predictions with the experimental result from LEP, we found the $Z$ invisible width to provide in general quite strong constraints.

\subsection{Electroweak Precision Observables} 

Majorana neutrinos will couple to the Z and W bosons, thus contributing to their self energies. The new physics contributions to electroweak radiative corrections are often parametrised through the Peskin-Tackeuchi parameters $S, T, U $: we have computed these observables and compared them with experimental observations. The resulting constraints are, in most scenarios considered here, weaker than those deriving from the LFV lepton decays and from the $Z$ invisible width.

\subsection{Neutrinoless double beta decay}

The hypothesis that neutrinos are Majorana fermions implies the possibility of lepton number violating processes, such as the the neutrinoless double beta: Majorana neutrinos are equal to their antineutrinos, so that two beta decays can combine, resulting in the decay of two nuclei with the emission of two electrons and no neutrino. This decay has not been observed yet by any of the high sensitivity experiments currently searching for it. In our ISS set-up the decay rate is dominated by the contribution of the light neutrinos, which ultimately results in a bound on the mass of the lightest neutrino, that is similar to the one obtained in the SM case.

\subsection{Theoretical consistency of the model}

We must require the total decay width of each heavy neutrino to be smaller than  its mass. In particular, we have applied the following condition:
\begin{equation}
\frac{\Gamma_{N_i}}{m_{N_i}}<\frac12 \quad {\rm for}~ i=1,\dots,6.
\label{widthconstraint}
\end{equation}

This results in a bound constraining the size of the global Yukawa coupling $f$ to be below order 2-3. It consequently restricts the entries of the Yukawa matrix to $|(Y_\nu)_{ij}|^2 /(4 \pi)<1/2-1$, and is therefore sufficient to ensure that we are working within the perturbative regime.

Finally, we must check that the predicted values of both the light neutrino mass squared differences and the neutrino mixing angles are actually in agreement with data. 
We have required the entries of the $U_\nu$ matrix belonging to the $ 3\times3 $ light neutrinos sub-block, obtained after the diagonalisation of the full neutrino mass matrix, to lie within the 3$\sigma$ experimental  bands. The same requirement is applied to the mass squared differences computed from the eigenvalues of the mass matrix that correspond to the light neutrinos.

\section{Maximum allowed Z decay rates within the ISS-$\cancel{\rm LFV}_{\mu e}$}\label{max Z decays}

\begin{figure}[p!]
\begin{center}
\includegraphics[width=.49\textwidth]{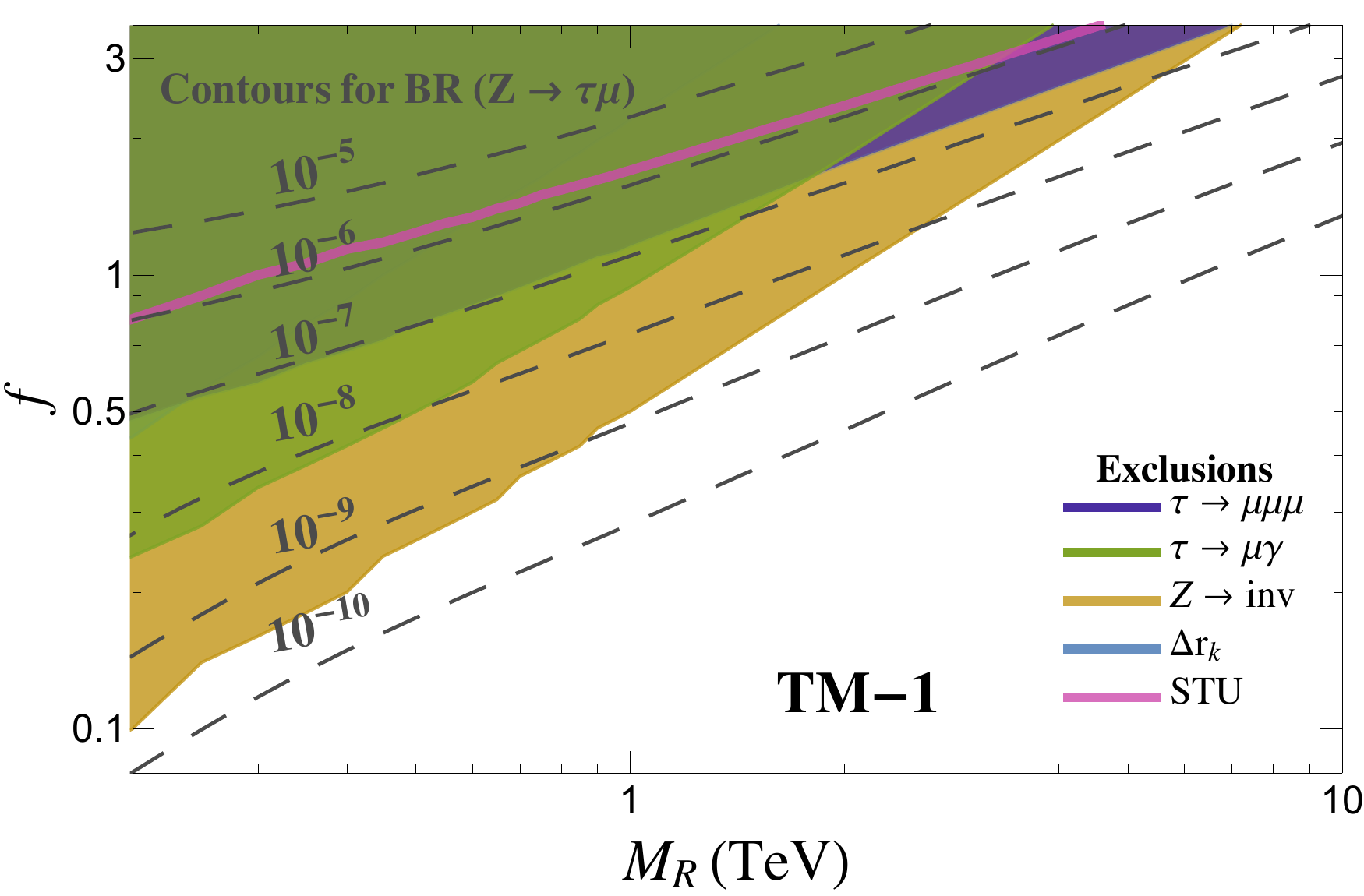}
\includegraphics[width=.49\textwidth]{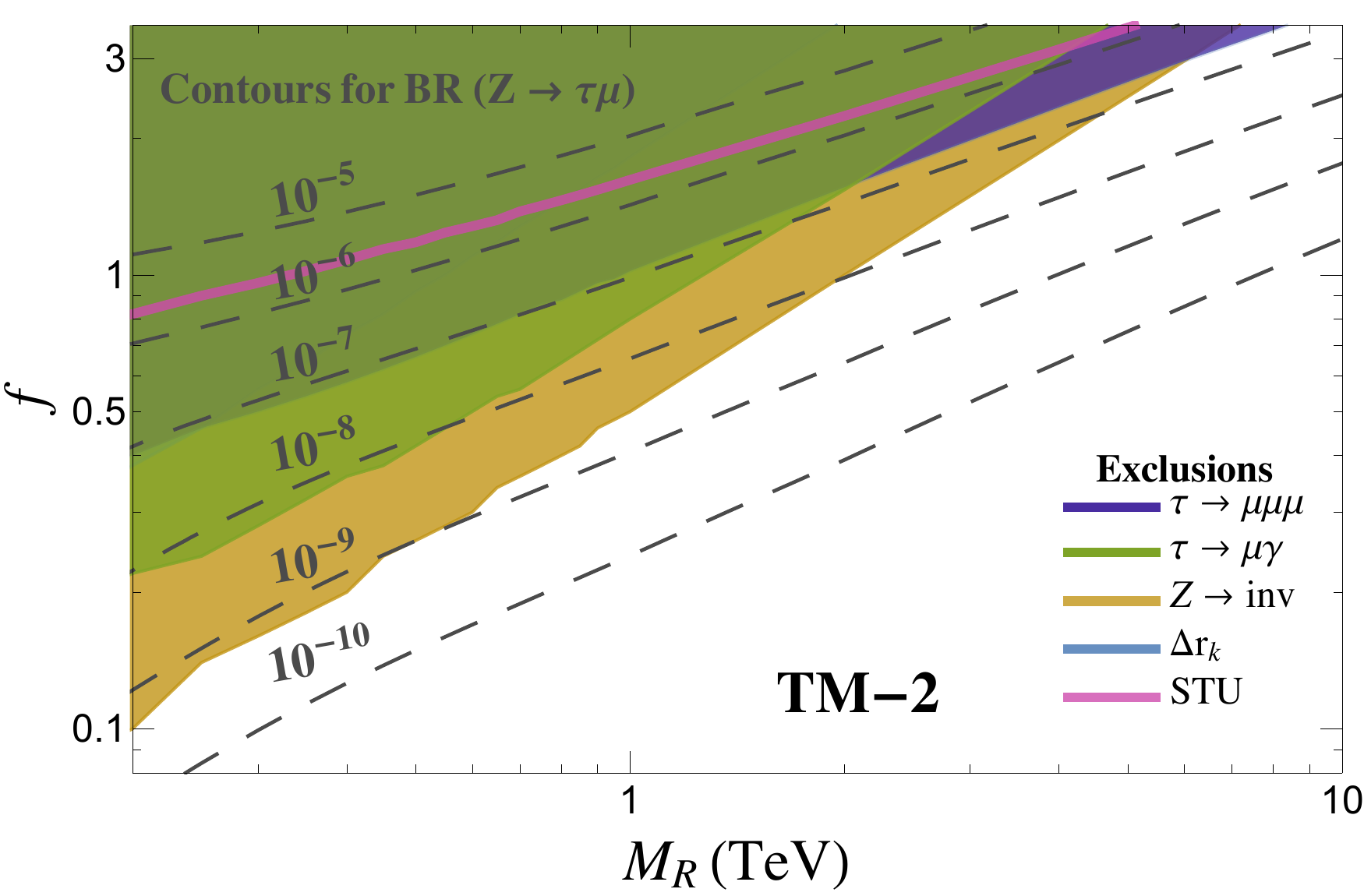}
\includegraphics[width=.49\textwidth]{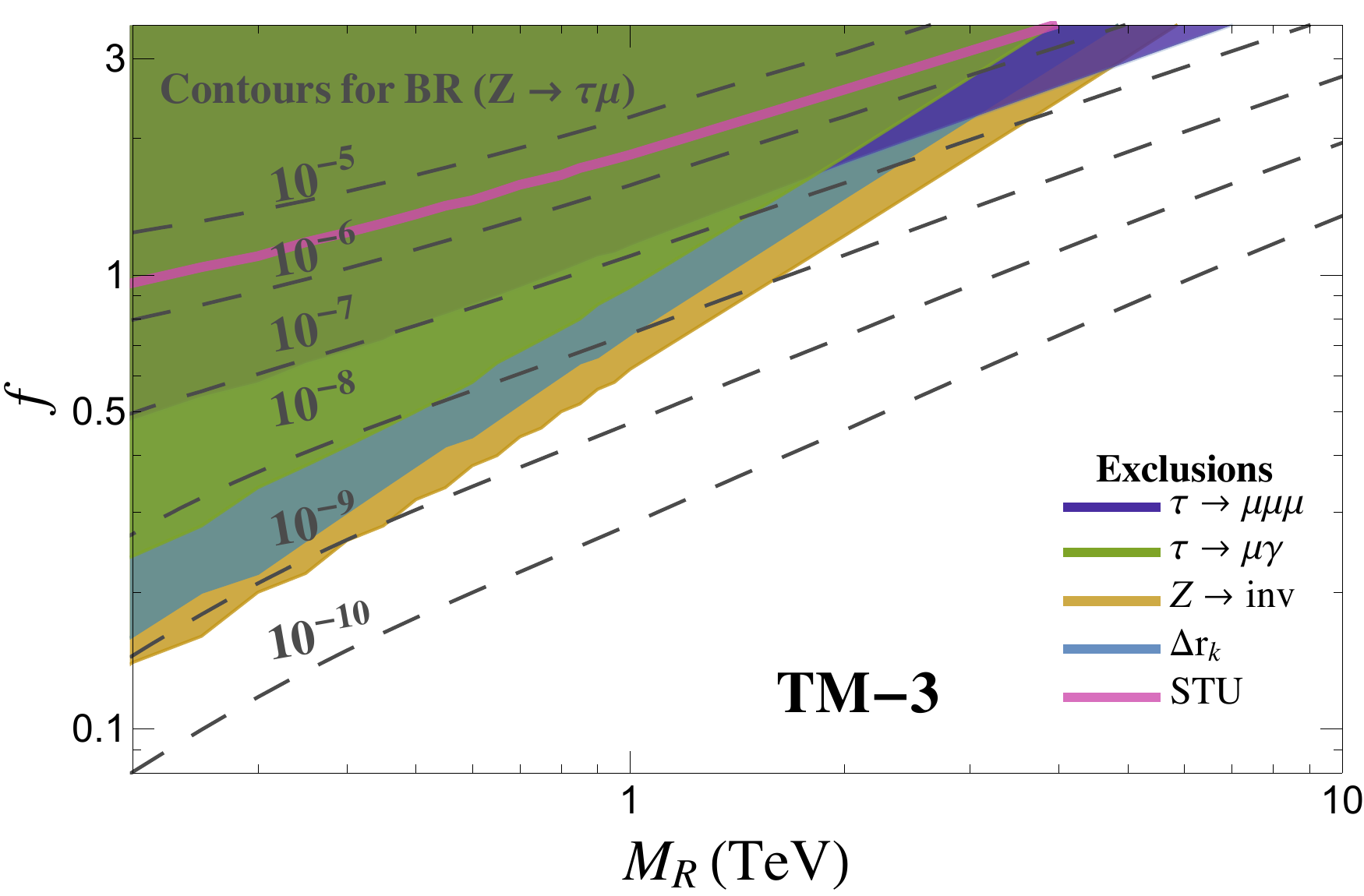}
\includegraphics[width=.49\textwidth]{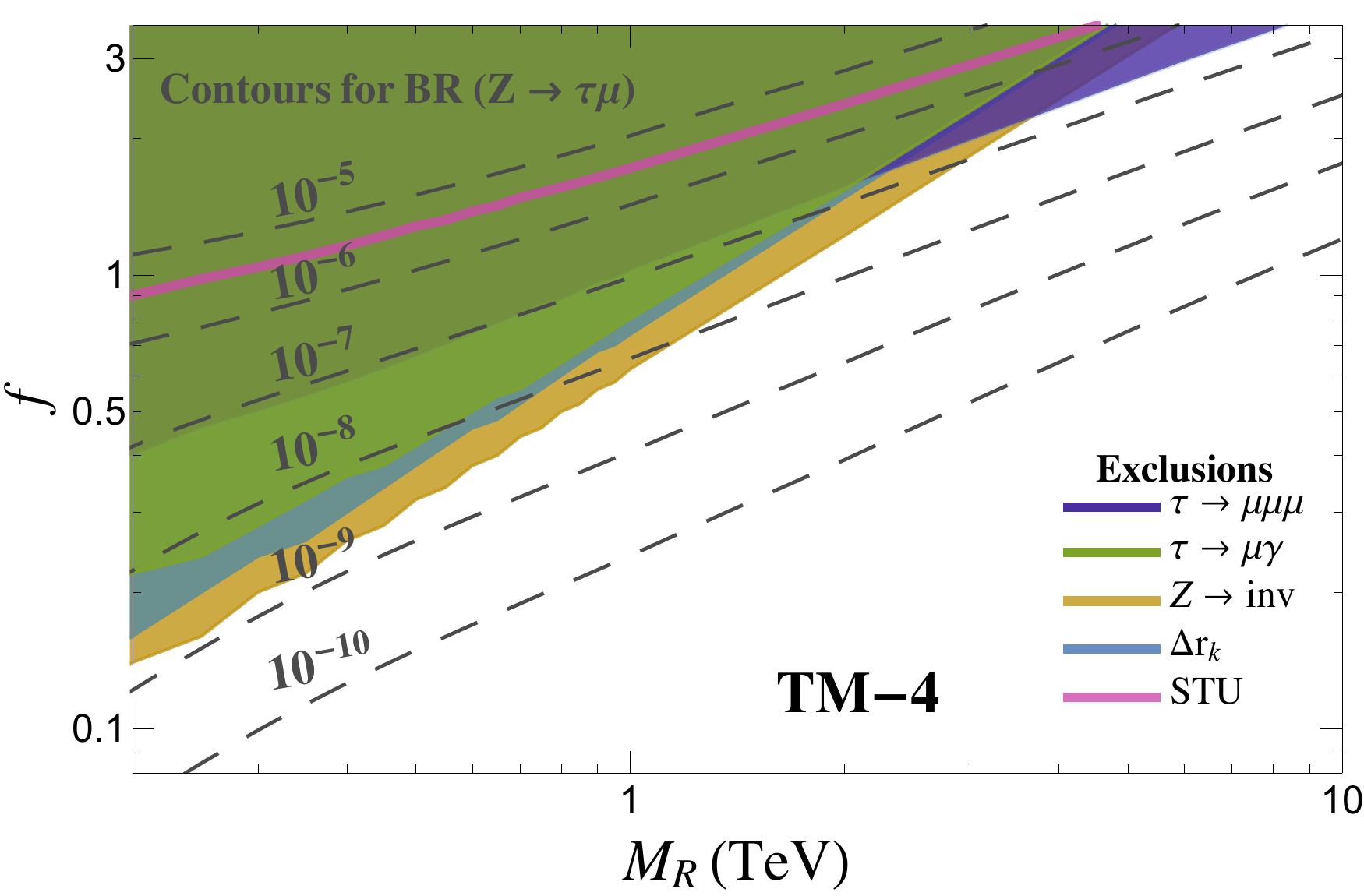}
\includegraphics[width=.49\textwidth]{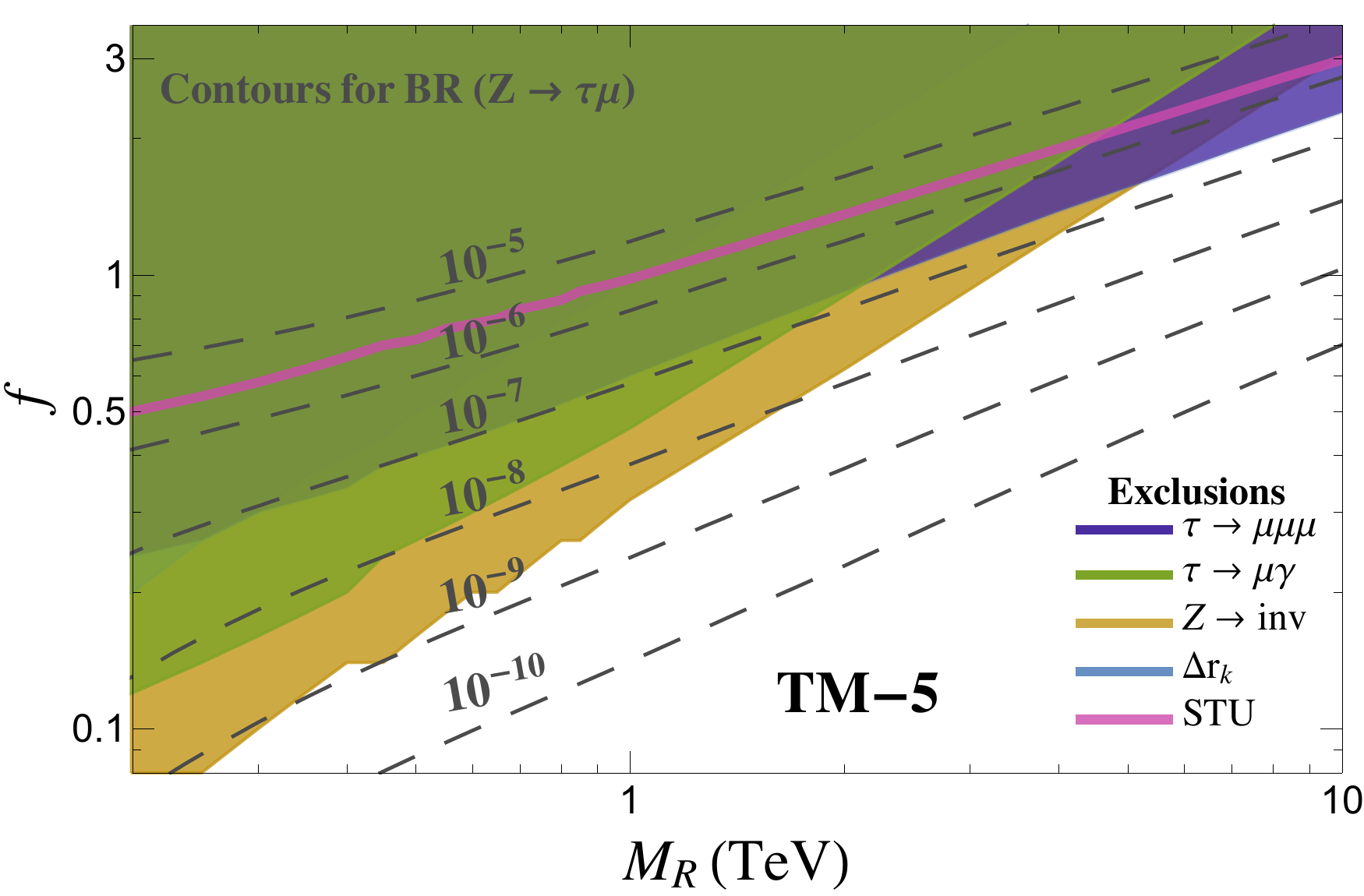}
\includegraphics[width=.49\textwidth]{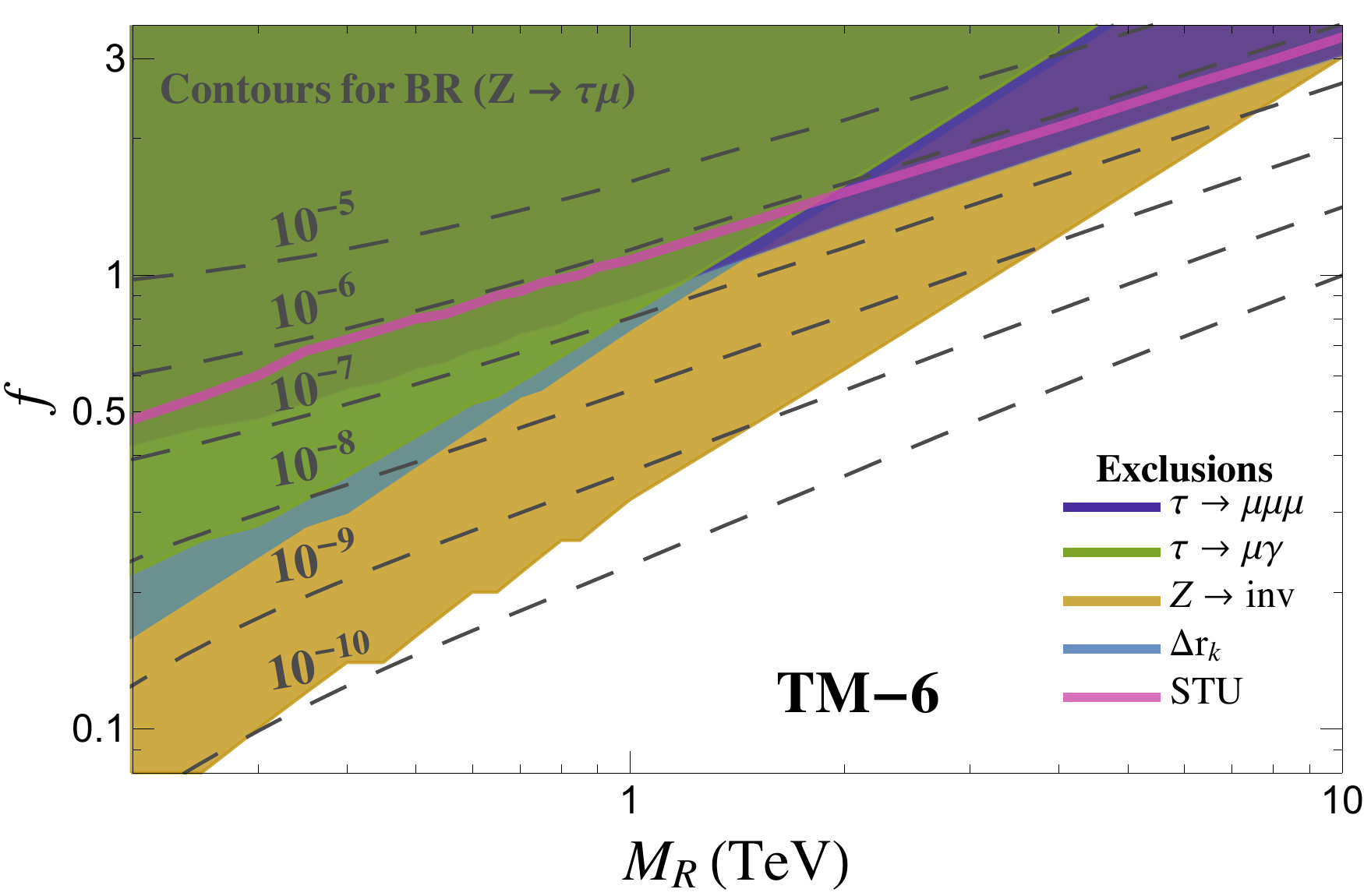}
\includegraphics[width=.49\textwidth]{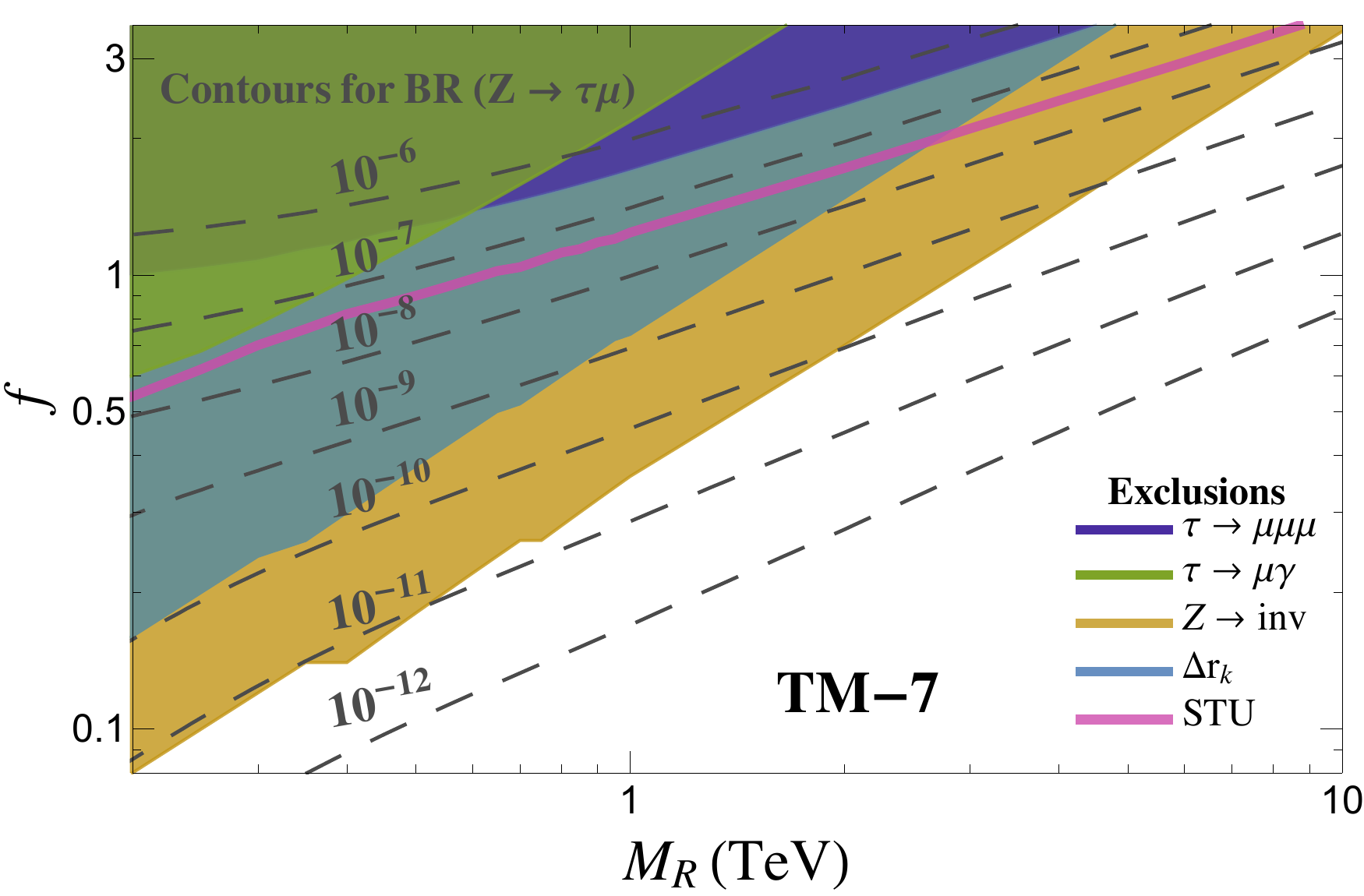}
\includegraphics[width=.49\textwidth]{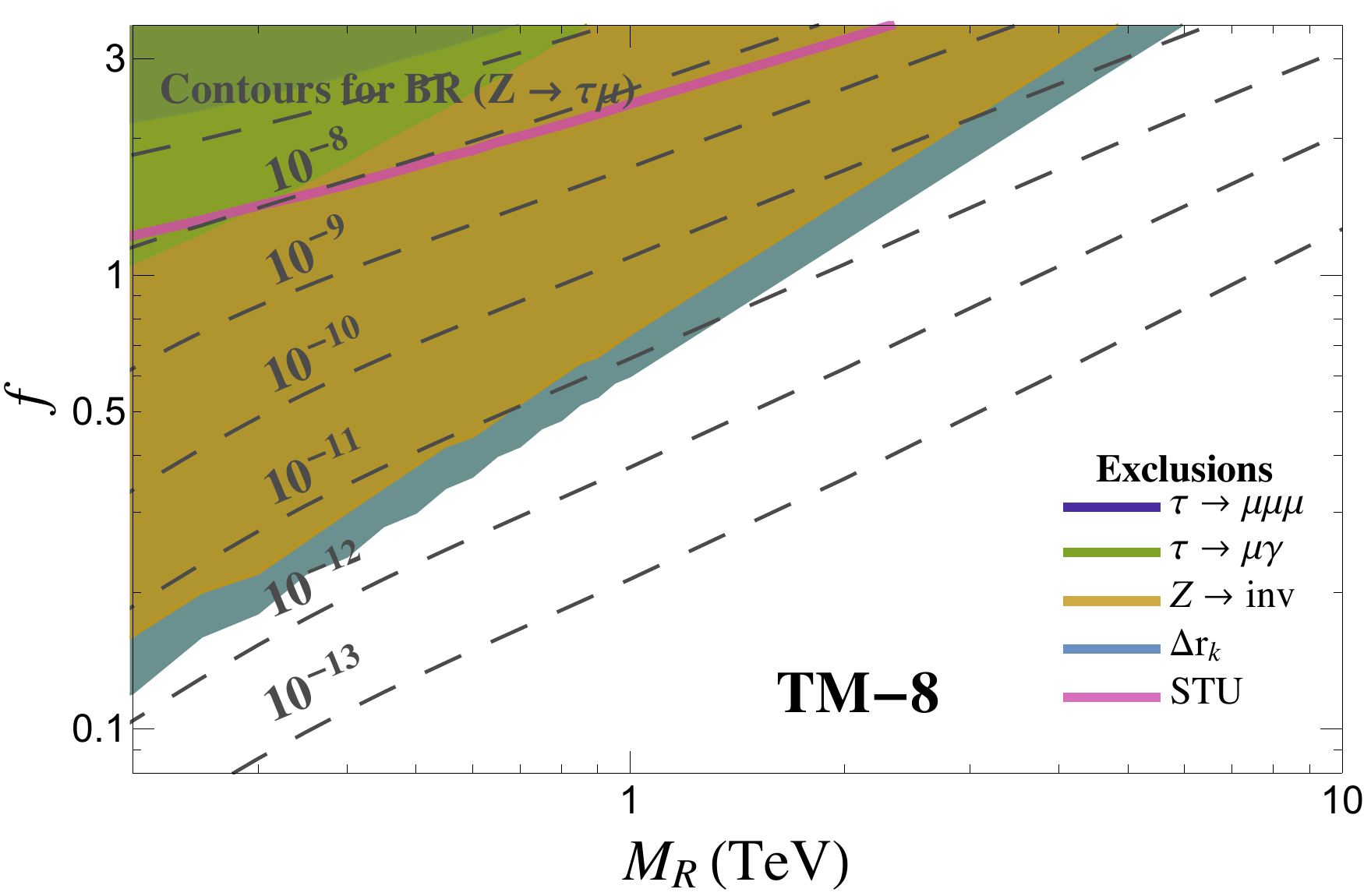}
\caption{Contour lines for BR($Z\to\tau\mu$) (dashed lines) in the ($M_R,f$) plane of the ISS model for the eight TM scenarios in Table \ref{TMscenarios}. Shadowed areas are the excluded regions by $\tau\to\mu\mu\mu$ (purple), $\tau\to\mu\gamma$ (green), Z invisible width (yellow) and  $\Delta r_k$ (cyan). The region above the pink solid line is excluded by the $S$, $T$, $U$ parameters. We obtain similar results for BR($Z\to\tau e$) in the TE scenarios by exchanging $\mu$ and $e$ in these plots of the TM scenarios.}\label{ZtaumufMRplane}
\end{center}
\end{figure}

To conclude, we have compared the predicted ratios for the LFVZD with the exclusions imposed by experiment for the eight sample Yukawa textures defined in Table \ref{TMscenarios}. Once we choose a given texture, the only remaining free parameters are the Yukawa strength $ f $ and the heavy mass scale $ M_R $. In Figure~\ref{ZtaumufMRplane} we plot in the ($M_R, f$) plane the exclusions resulting from the most relevant constraints and superpose the contour lines of the $Z\to\tau\mu$ branching ratio\footnote{We have verified that the parameter space allowed by experimental constraints is also theoretically consistent.}.
Notice how the contour corresponding to BR$(Z\to\tau\mu)=2 \times 10^{-7}$ is almost superposed to the exclusion bound from the three body decay, in agreement with the correlation mentioned above.
In the large $M_R$ region this is the most constraining observable: BR$(Z\to\tau\mu)$ as large as $2 \times 10^{-7}$ are allowed for high values of $M_R$, in most sample textures. Exceptions are the textures TM-7 and TM-8, where the small values of $ \ctm $ and $\modmu$, $\modtau$ respectively result in suppressed $LFV_{\mu\tau}$ transitions. The $Z$ invisible width is instead typically the most constraining observable in the region of low $M_R$ values. The maximum allowed LFV $Z$ decay rates BR$(Z\to\tau\mu)\sim 2 \times 10^{-7}$ are then obtained in the crossing region of these two excluded areas. This crossing occurs for different values of the heavy mass scale $M_R$ in each scenario, being as low as 2-4 TeV  for cases TM-4 and TM-5.

\section{Conclusions}
\label{conclusions}
In this work we have studied several aspects of cLFV processes in the context of the ISS model. 
Motivated by the strong experimental upper bounds on LFV $\mu$-$e$ transitions, we have discussed a useful geometrical parametrisation of the neutrino Yukawa coupling matrix that allows us to easily define ISS scenarios with suppressed ${\rm LFV}_{\mu e}$.
We have studied in full detail the LFV $Z$ decays in these scenarios as well as the observables setting the most relevant experimental constraints on the model's parameter space.
Taking into account all the relevant bounds, we found that heavy ISS neutrinos with masses in the few TeV range can induce rates up to BR$(Z\to\tau\mu)\sim 2 \times 10^{-7}$ and BR$(Z\to\tau e)\sim 2 \times 10^{-7}$ in the TM and TE scenarios, respectively.
These rates are potentially measurable at future linear colliders and FCC-ee. \\
Thus, we have shown that searches for LFVZD at future colliders may be a powerful tool to probe cLFV in the context of low scale Seesaw models. Furthermore, the large LFVZD rates here predicted could  be induced by heavy neutrinos lying in a mass range that allows direct production at LHC. Full details of this work and a complete list of references can be found in Ref.~\cite{DeRomeri:2016gum}.

\section*{Acknowledgements}
F. S. wishes to express her gratitude to the organizers of the Corfu Summer Institute 2016 "School and Workshops on Elementary Particle Physics and Gravity" for giving her the opportunity to present this study. This work has been supported by the European Union through the ITN ELUSIVES H2020-MSCA-ITN-2015//674896 and the RISE INVISIBLESPLUS H2020-MSCA-RISE-2015//690575, by the Spanish AEI (Agencia Estatal de Investigacion) and the EU FEDER (Fondo Europeo de Desarrollo Regional) through the project FPA2016-78645-P, by the Spanish Consolider-Ingenio 2010 Programme CPAN (CSD2007-00042) and by the Spanish MINECO's ``Centro de Excelencia Severo Ochoa''  Programme under grant SEV-2012-0249. X. M. is supported through the FPU grant AP-2012-6708.

\bibliography{bibliography}

\begin{thebibliography}{99}



\bibitem{DeRomeri:2016gum} 
  V.~De Romeri, M.~J.~Herrero, X.~Marcano and F.~Scarcella,
  Phys.\ Rev.\ D {\bf 95}, no. 7, 075028 (2017)
  doi:10.1103/PhysRevD.95.075028
  [arXiv:1607.05257 [hep-ph]].
  
  
\bibitem{Minkowski:1977sc} 
  P.~Minkowski,
  Phys.\ Lett.\  {\bf 67B}, 421 (1977).
  doi:10.1016/0370-2693(77)90435-X




\bibitem{Bernabeu:1987gr} 
  J.~Bernabeu, A.~Santamaria, J.~Vidal, A.~Mendez and J.~W.~F.~Valle,
  Phys.\ Lett.\ B {\bf 187}, 303 (1987).
  doi:10.1016/0370-2693(87)91100-2



\bibitem{Abada:2014cca} 
  A.~Abada, V.~De Romeri, S.~Monteil, J.~Orloff and A.~M.~Teixeira,
  JHEP {\bf 1504}, 051 (2015)
  doi:10.1007/JHEP04(2015)051
  [arXiv:1412.6322 [hep-ph]].


\bibitem{Abada:2015zea} 
  A.~Abada, D.~Be\v{c}irevi\'{c}, M.~Lucente and O.~Sumensari,
  Phys.\ Rev.\ D {\bf 91}, no. 11, 113013 (2015)
  doi:10.1103/PhysRevD.91.113013
  [arXiv:1503.04159 [hep-ph]].




\bibitem{Arganda:2014dta} 
  E.~Arganda, M.~J.~Herrero, X.~Marcano and C.~Weiland,
  Phys.\ Rev.\ D {\bf 91}, no. 1, 015001 (2015)
  doi:10.1103/PhysRevD.91.015001
  [arXiv:1405.4300 [hep-ph]].




\end{thebibliography}

\end{document}